\documentclass[12pt,a4wide]{JHEP3}
\usepackage{amsmath,amssymb}
\usepackage{braket}
\usepackage{graphicx}

\title{The particle-hole transformation, supersymmetry and achiral boundaries of the open Hubbard model}

\author{Alejandro De La Rosa Gomez\thanks{Email: alrg500@york.ac.uk}\\
Department of Mathematics, University of York, York YO10 5DD, UK
}

\abstract{We show that the particle-hole transformation in the Hubbard model has a crucial role in relating Shastry's $R$-matrix to the AdS/CFT $S$-matrix. In addition, we construct an achiral boundary for the open Hubbard chain which possesses twisted Yangian symmetry.}
\begin{document}

\begin{section}{Introduction}
The Hubbard model \cite{Hubbard} is one of the classic integrable models of condensed matter physics, especially useful in describing the transition from conducting to insulating systems and vice versa.  Its symmetry is of Yangian type \cite{Korepin}, but is insufficient to fix its $R$-matrix \cite {Shastry}, which possesses a very atypical structure -- it cannot be written as a function of the difference of the spectral parameters.

This model has aroused a lot of interest recently due to its connections with integrable systems in the worldsheet scattering picture of the AdS/CFT correspondence. Remarkable results \cite{Beisert,Mitev,Vidas2} include the identification of the Hubbard model $R$-matrix with the centrally extended $\mathfrak{su}$(2\textbar 2) (or `AdS/CFT') $S$-matrix, and the connection between the closed and open $q$-deformed Hubbard chain and the $U_q(\mathfrak{su}$(2\textbar 2)) spin chain. This has been shown to be a consequence of the Hubbard model's symmetry being the bosonic subalgebra of $\mathfrak{su}$(2\textbar 2), which may be enhanced to the full superalgebra when certain conditions on the AdS/CFT variables are satisfied \cite{Beisert3}.

In this paper we explain how the key to this enhancement is the generator of the particle-hole transformation (PHT), which relates the left and right $\mathfrak{su}(2)$ symmetries of the model. Although the PHT is neither a symmetry of the hamiltonian nor the $R$-matrix, it does allow for a more specific solution of Shastry's ansatz to be invariant under the full superalgebra. In addition, this transformation provides us with a boundary theory that possesses a twisted Yangian, further improving our knowledge of the integrable structure governing particle scattering in the Hubbard model.

This paper is organized as follows. First, we review the symmetry algebra of the Hubbard model and its Yangian extension. Secondly, we show the role of the particle-hole transformation in upgrading the bosonic $\mathfrak{su}(2)^2$ to $\mathfrak{su}$(2\textbar 2) and how it can be identified a a linear combination of the supercharges in $\mathfrak{su}$(2\textbar 2). In addition, we will show it is possible to obtain the AdS/CFT $S$-matrix from Shastry's $R$-matrix, even when the sign of the coupling constant changes under the particle-hole transformation. Then, we will review briefly the theory of twisted Yangians in the presence of achiral boundaries -- an algebraic structure which also appears in integrable boundary theories in the AdS/CFT correspondence \cite{vidas3}. Finally, we will construct the twisted Yangian symmetry for a half-infinite Hubbard chain with a boundary that reflects a particle as a hole, proving two crucial properties: that it commutes with the Hamiltonian and that it forms a coideal subalgebra of the original Yangian. The details of these calculations, together with the $\mathfrak{su}$(2\textbar 2)$\ltimes\mathbb{R}^2$ defining relations, are presented in an appendix.  

\end{section}

\begin{section}{Symmetries of the Hubbard Model}
The Hubbard model \cite{Hubbard} is an approximate theory used in solid state physics to describe how interactions between electrons can give rise to conducting and insulating systems. It is a spin chain with $N$ sites with Hamiltonian
\begin{equation}
H=-\sum_{i=1}^N\sum_{\sigma= \uparrow,\downarrow} c_{i\sigma}^{\dagger}c_{i+1\sigma}+c_{i+1\sigma}^{\dagger}c_{i\sigma}+U\sum_{i=1}^N(n_{i\uparrow}-\frac{1}{2})(n_{i\downarrow}-\frac{1}{2})
\end{equation}
where $c_{i\sigma}^{\dagger}$ and $c_{i\sigma}$ are the usual fermionic creation and annihilation operators acting on site $i$ and satisfying the only nonvanishing anticommutation relation
\begin{equation}
\{c_{i\sigma}^{\dagger}c_{j\tau}\}=\delta_{\sigma\tau}\delta_{ij} \ , 
\end{equation}
$U$ is the coupling constant for the on-site interaction and $n_{i\sigma}=c_{i\sigma}^{\dagger}c_{i\sigma}$ is the number density operator. There are four fundamental states per spin site, two bosonic $\ket{\phi^a}$ and two fermionic $\ket{\psi^{\alpha}}$:
\begin{equation}
\ket{\phi^1}= \ket{0}, \quad \ket{\psi^1} = c^{\dagger}_{\downarrow}\ket{0}, \quad \ket{\psi^2} = c^{\dagger}_{\uparrow}\ket{0}, \quad \ket{\phi^2} = c^{\dagger}_{\uparrow}c^{\dagger}_{\downarrow}\ket{0}.
\end{equation}
This model was shown to be quantum integrable when imposing both periodic and open boundary conditions \cite{Hubbard, SW3}, the latter leading to a twisted Yangian symmetry \cite{AM}. Its $R$-matrix, which we shall denote as $R$, can be identified as a linear combination of tensor products of two free fermion model $R$-matrices, one for each spin layer \cite{SW}. $R$ is a function of $U$ and the rapidity at site $k$, $\theta_k$, and these are usually grouped in functions $a_k=a(\theta_k)$, $b_k=b(\theta_k)$ and $h_k = h(U,\theta_k)$. Interestingly, one can relate $R$ to the $S$-matrix of the $AdS_5 \times S^5$ superstring, which possesses $\mathfrak{su}$(2\textbar2)$\ltimes \mathbb{R}^2$ symmetry, via a similarity transformation. It then is not surprising how much interest this model and its quantum deformation have recently aroused in the study of integrable systems in the AdS/CFT correspondence. 
The key to these connections is the $\mathfrak{su}(2)_L\times \mathfrak{su}(2)_R \subset \mathfrak{su}$(2\textbar 2) symmetry of the model \cite{Beisert}, with the left and right copies of $\mathfrak{su}(2)$ acting on the fermionic and bosonic states respectively. However, this does not explain how the scattering picture of a nonsupersymmetric model -- unlike extensions of (2.1) \cite{DFFR} -- can lead to one of a supersymmetric theory, especially when the $R$-matrix must commute with the existing supercharges. Indeed, one needs an additional constraint to accomplish this \cite{Beisert3}. Specifically, if the AdS/CFT variables $x^-$, $x^+$ and $g$ are identified with the Hubbard variables as follows
\begin{equation}
g = \frac{1}{U}, \quad x^+ = \frac{ib}{aU}e^{2h}, \quad x^- = \frac{a}{ibU}e^{2h},
\end{equation}
then $\mathfrak{su}(2)_L\times \mathfrak{su}(2)_R$ can be enhanced to $\mathfrak{su}$(2\textbar 2) if
\begin{equation}
\left(\frac{x^+}{x^-}\right)^{L/2}=1.
\end{equation}
where $L$ is the length of the chain. This condition is derived using the Bethe ansatz method for a spin chain with $\mathfrak{su}$(2\textbar 2)$\ltimes \mathbb{R}^2$ symmetry before identifying variables according to (2.4). Clearly, there must also be an explanation for this condition coming from the Hubbard model itself. This requires a slightly deeper study of the full symmetry algebra of the Hubbard model and the intertwiner which relates the two copies of $\mathfrak{su}(2)$. If one defines the following operators:
\begin{equation}
\mathcal{E}_{iL}^n=c_{i\uparrow}^{\dagger}c_{i+n\downarrow}, \quad \mathcal{F}_{iL}^n=c_{i\downarrow}^{\dagger}c_{i+n\uparrow}, \quad \mathcal{H}_{iL}^n=c_{i\uparrow}^{\dagger}c_{i+n\uparrow}-c_{i\downarrow}^{\dagger}c_{i+n\downarrow}.
\end{equation}
where $i$ is the spin chain site and $n \in \mathbb{Z}$, then $\mathfrak{su}(2)_L$ is generated by $\{E_{0L},F_{0L},H_{0L}\}$
\begin{equation}
E_{0L}=\frac{1}{\sqrt{2}}\sum_i\mathcal{E}^0_{iL}, \quad F_{0L}=\frac{1}{\sqrt{2}}\sum_i\mathcal{F}^0_{iL}, \quad H_{0L}=\frac{1}{2}\sum_i\mathcal{H}^0_{iL},
\end{equation}
where $i$ runs over all possible spin chain sites and the operators satisfy $[H_{0L},E_{0L}]=E_{0L},\ [H_{0L},F_{0L}]=-F_{0L}$ and $[E_{0L},F_{0L}]=H_{0L}$. The $\mathfrak{su}(2)_R$ algebra, also known as the eta-pairing symmetry \cite{SW2}, is generated by $\{E_{0R},F_{0R},H_{0R}\}$, which can be obtained through the \emph{``partial" particle hole transformation} $P_{\downarrow}$:
\begin{equation}
P_{\downarrow}: \quad (c_{i \downarrow},c_{i \downarrow}^{\dagger},c_{i \uparrow},c_{i \uparrow}^{\dagger}) \ \mapsto \ ((-1)^{i}c_{i \downarrow}^{\dagger},(-1)^{i}c_{i \downarrow},c_{i \uparrow},c_{i \uparrow}^{\dagger})
\end{equation}
Similarly, one can also obtain the generators of $\mathfrak{su}(2)_R$ via the following equivalent map, which we shall denote by $P_{\uparrow}$:
\begin{equation}
P_{\uparrow}: \quad (c_{i \downarrow},c_{i \downarrow}^{\dagger},c_{i \uparrow},c_{i \uparrow}^{\dagger}) \ \mapsto \ (c_{i \downarrow},c_{i \downarrow}^{\dagger},(-1)^{i}c_{i \uparrow}^{\dagger},(-1)^{i}c_{i \uparrow})
\end{equation}
One must note that the eta-pairing symmetry is only present at the global level if the length of the chain is even \cite{Beisert3}. Similarly, $\mathfrak{su}$(2\textbar 2)$\ltimes \mathbb{R}^2$ is only present locally -- a mismatch of phases in the spin chain will break it globally.
As expected from integrability, the model also possesses a Yangian symmetry \cite{Korepin}, which includes the original -- or grade 0 -- generators of the algebra and a second set of generators in the vector representation -- the grade 1 generators. This Yangian was constructed \cite{Korepin} for $N \rightarrow  \infty$ and, as expected, it is composed of two copies of $Y(\mathfrak{su}(2))$ related by $P_{\sigma}$, $\sigma=\downarrow,\uparrow$. $Y(\mathfrak{su}(2)_L)$ is generated by $\{E_{kL},F_{kL},H_{kL}\}_{k=0,1}$, where the grade 1 generators are given by 
\begin{eqnarray}
E_{1L}&=&\frac{1}{\sqrt{2}}\sum_i (\mathcal{E}_{iL}^1-\mathcal{E}_{iL}^{-1})-\frac{U}{2\sqrt{2}}\sum_{i<j}(\mathcal{E}_{iL}^0\mathcal{H}_{jL}^0-\mathcal{E}_{jL}^0\mathcal{H}_{iL}^0), \nonumber \\
F_{1L}&=&\frac{1}{\sqrt{2}}\sum_i (\mathcal{F}_{iL}^1-\mathcal{F}_{iL}^{-1})+\frac{U}{2\sqrt{2}}\sum_{i<j}(\mathcal{F}_{iL}^0\mathcal{H}_{jL}^0-\mathcal{F}_{jL}^0\mathcal{H}_{iL}^0), \nonumber \\
H_{1L}&=&\frac{1}{2}\sum_i (\mathcal{H}_{iL}^1-\mathcal{H}_{iL}^{-1})+\frac{U}{2}\sum_{i<j}(\mathcal{E}_{iL}^0\mathcal{F}_{jL}^0-\mathcal{E}_{jL}^0\mathcal{F}_{iL}^0).
\end{eqnarray}
$P_{\sigma}$ is a map between representations of the same algebra, but it is not necessarily a symmetry of the theory. In the case of the Hubbard model, however, both the hamiltonian and the fermionic $R$-matrix $R_f$ \cite{SW2} satisfy
\begin{equation}
P_{\sigma} Z(\theta,U) \mapsto Z(\theta,-U), \quad \quad Z=H, R_f.
\end{equation}
and hence the map $P_{\sigma}$ combined with a change of sign in $U$ is an additional symmetry of the model -- more specifically, a supersymmetry. 

The $R$-matrix $R$ from which one can obtain the AdS/CFT $S$-matrix is related to $R_f$ in the following way:
\begin{equation}
(R_f)_{12} = W^{-1}R_{12}(a_1,a_2,b_1,b_2)W \lvert_{a_j=\cos(\theta_j),b_j=-i\sin(\theta_j)}
\end{equation}
where $W$ is the matrix:
\begin{equation}
W = \textrm{diag}(1,1,-i,-i,-i,-i,1,1,-1,-1,i,i,i,i,-1,-1)
\end{equation} 
and $R$ for such values of $a_j$ and $b_j$ corresponds to Shastry's $R$-matrix \cite{Shastry}. Here we will consider $R$ for general $a_j$ and $b_j$. Since the Hubbard model can be interpreted as two coupled XX models -- each one corresponding to a different spin -- $R$ can be written as a linear combination of tensor products of two types of XX-model $R$-matrices $R_{12 \sigma}^{\pm}$ \cite{Mitev, SW}:
\begin{equation}
R_{12}=A_{12}(R_{12 \uparrow}^+ \otimes R_{12 \downarrow}^+ + R_{12 \uparrow}^- \otimes R_{12 \downarrow}^-) + R_{12 \uparrow}^+ \otimes R_{12 \downarrow}^- + R_{12 \uparrow}^-\otimes R_{12 \downarrow}^+,
\end{equation}
where 
\begin{equation}
A_{12} = \frac{-\frac{b_1}{a_1}\sqrt{1+\frac{U}{4}(a_1 b_1)^2}+\frac{a_2}{b_2}\sqrt{1+\frac{U}{4}(a_2 b_2)^2}-\frac{U}{2}(b_1^2+a_2^2)}{-\frac{b_2}{a_2}\sqrt{1+\frac{U}{4}(a_2 b_2)^2}+\frac{a_1}{b_1}\sqrt{1+\frac{U}{4}(a_1 b_1)^2}-\frac{U}{2}(b_2^2+a_1^2)}\left(\frac{a_1b_2}{a_2b_1}\right).
\end{equation}
Since this $R$-matrix can be related to the AdS/CFT $S$-matrix via a similarity transformation, we will show it to be supersymmetric under an identification among its variables given by the PHT.

Now we shall proceed to use (2.8), (2.9) and (2.11) to construct the form of $P_{\sigma}$. Then we will check that this supersymmetry is enough to enhance the bosonic symmetry of the model to $\mathfrak{su}$(2\textbar 2). In addition, we will show that this enhancement imposes a condition in $R$ equivalent to (2.5), which makes the $U$-dependence disappear. This explains why, in this case, $R$ is supersymmetric though the Hubbard model is not.

\end{section}

\begin{section}{The Particle Hole Transformation: from $\mathfrak{su}(2)^2$ to $\mathfrak{su}$(2\textbar 2)$\ltimes \mathbb{R}^2$}
Recall the standard 2 $\times$ 2 $\mathfrak{su}$(2)-triple $\{\frac{1}{\sqrt{2}}e,\frac{1}{\sqrt{2}}f,\frac{1}{2}h\}$, where 
\begin{equation}
e =  \begin{pmatrix}
0&1\\0&0
\end{pmatrix}, \quad
f = \begin{pmatrix}
0&0\\1&0
\end{pmatrix}, \quad
h =  \begin{pmatrix}
1&0\\0&-1
\end{pmatrix}.
\end{equation}
If one computes matrices $M$ which commute with the fermionic $R$-matrix as given in \cite{SW2},
\begin{equation}
(M \otimes M) R_f = R_f (M \otimes M),
\end{equation}
one obtains\begin{equation}
M = 
\begin{pmatrix}
R_1 &0&0&R_2\\0&L_1&L_2&0\\0&L_3&L_4&0\\R_3&0&0&R_4
\end{pmatrix}
\end{equation}
where the entries satisfy $R_1R_4-R_2R_3=L_1L_4-L_2L_3=\Delta$. We can set $\Delta =1$ since it does not affect integrability, and we have:\begin{equation}
M_L=\begin{pmatrix}
L_1&L_2\\L_3&L_4
\end{pmatrix} \in SU(2)_L \quad M_R=\begin{pmatrix}
R_1&R_2\\R_3&R_4 
\end{pmatrix} \in SU(2)_R.
\end{equation}
Then one obtains a representation of the $\mathfrak{su}(2)$ algebra given by $\mathfrak{su}(2)_L = \{E_{0L},F_{0L}, H_{0L}\} = \{\frac{1}{\sqrt{2}}(e \otimes f),\frac{1}{\sqrt{2}}( f \otimes e), \frac{1}{4}(h \otimes 1 - 1 \otimes h)\}$ and an additional, commuting copy given by $\mathfrak{su}(2)_R = \{E_{0R},F_{0R}, H_{0R}\} = \{\frac{1}{\sqrt{2}}(e \otimes e), \frac{1}{\sqrt{2}}(f \otimes f), \frac{1}{4}(h \otimes 1 + 1 \otimes h)\}$. Now one finds that the partial particle hole transformations, which map the copies of $\mathfrak{su}(2)$ to one another and satisfies (2.11), are each divided into two possible choices:
\begin{eqnarray}
P_{\downarrow}^{\pm}(\textbf{a}_1) &=& \textbf{a}_1 (1 \otimes (e\pm f)),\\
P_{\uparrow}^{\pm}(\textbf{a}_2) &=& \textbf{a}_2 ((e\pm f) \otimes 1),
\end{eqnarray}
where \textbf{a}$_1$ and \textbf{a}$_2$ are nonzero complex numbers. We shall proceed to relate these to the supersymmetry charges $\mathbb{Q}$ and $\mathbb{G}$ of $\mathfrak{su}$(2\textbar 2)$\ltimes \mathbb{R}^2$. Using the representation of the bosonic subalgebra given above, and the defining relations in the Appendix A.1, the supercharges are
\begin{align}
&\mathbb{Q}^1_{\ 1}(\textbf{a}, \textbf{b})= (\textbf{b} ef+\textbf{a} fe)\otimes f,
&&\mathbb{Q}^2_{\ 2}(\textbf{a}, \textbf{b})= (\textbf{a} ef+\textbf{b} fe)\otimes e,& \nonumber\\
&\mathbb{Q}^1_{\ 2}(\textbf{a},\textbf{b}) = e \otimes (\textbf{a} ef - \textbf{b} fe),&&
\mathbb{Q}^2_{\ 1}(\textbf{a},\textbf{b}) = -f \otimes (\textbf{b} ef - \textbf{a} fe),& \nonumber \\
&\mathbb{G}^1_{\ 1}(\textbf{c},\textbf{d}) = \mathbb{Q}^2_{\ 2}(\textbf{c},\textbf{d}), &&
\mathbb{G}^2_{\ 2}(\textbf{c},\textbf{d}) = \mathbb{Q}^1_{\ 1}(\textbf{c},\textbf{d}),& \nonumber\\
&\mathbb{G}^1_{\ 2}(\textbf{c},\textbf{d}) =- \mathbb{Q}^1_{\ 2}(\textbf{c},\textbf{d}), &&
\mathbb{G}^2_{\ 1}(\textbf{c},\textbf{d}) = -\mathbb{Q}^2_{\ 1}(\textbf{c},\textbf{d}), &
\end{align}
where the bold variables are nonzero complex numbers satisfying \textbf{ad} $-$ \textbf{bc} $=1$. It is now easy to see that the operators $P_{\sigma}^{\pm}$ are sums of these supercharges with a specific choice of variables:
\begin{eqnarray}
P_{\downarrow}^{\pm}(\textbf{a}) &=& \mathbb{Q}^1_{ \ 1}(\textbf{a,$\pm$a})+\mathbb{Q}^2_{ \ 2}(\textbf{a,$\pm$a})=\mathbb{G}^2_{ \ 2}(\textbf{a,$\pm$a})+\mathbb{G}^2_{ \ 2}(\textbf{a,$\pm$a}),\\
P_{\uparrow}^{\pm}(\textbf{c}) &=& \mathbb{Q}^1_{ \ 2}(\textbf{c,$\mp$c})+\mathbb{Q}^2_{ \ 1}(\textbf{c,$\mp$c})=\mathbb{G}^1_{ \ 2}(\textbf{c,$\mp$c})+\mathbb{G}^2_{ \ 1}(\textbf{c,$\mp$c}),
\end{eqnarray}
and hence the supercharges can be obtained by computing commutators of the particle hole transformation with the generators of the bosonic subalgebra. In this case however, imposing the condition \textbf{ad} $-$ \textbf{bc} = 1 is equivalent to the following relation among the free parameters:
\begin{equation}
\textbf{c} = \pm\frac{1}{2\textbf{a}}.
\end{equation}
Consequently, the superalgebra generated by $\mathfrak{su}(2)_L \times \mathfrak{su}(2)_R$ and $P_{\sigma}^{\pm}$ is $\mathfrak{su}$(2\textbar 2). This symmetry lacks the central extension that governs the scattering of the $AdS_5 \times S^5$ superstring. Instead, the possible central charges $\mathbb{C}$, $\mathbb{P}$ and $\mathbb{K}$ generated by the supersymmetries (see A.1) are
\begin{equation}
\langle \mathbb{C},\mathbb{P},\mathbb{K} \rangle = \langle \  \frac{\textbf{ad}+\textbf{bc}}{2}, \ \textbf{ab}, \ \textbf{cd} \ \rangle = \langle \ 0, \mp\textbf{a}^2, \pm \frac{1}{4\textbf{a}^2} \ \rangle.
\end{equation}
We can see that the relations \textbf{b} = $\mp$ \textbf{a} and \textbf{d} = $\pm$ \textbf{c} are equivalent to condition (2.5), which is ultimately due to the existence of the particle-hole transformation. 

If one now takes $R$ as given in (2.13) and imposes condition $b_j = \pm a_j$, one obtains 
\begin{equation}
R_{12}(b_j = \pm a_j) = (R_{12 \uparrow}^+ + R_{12 \uparrow}^-) \otimes  (R_{12 \downarrow}^+ + R_{12 \downarrow}^-),
\end{equation}
which is now invariant under the change of sign in $U$, and hence possesses $P^{\pm}_{\sigma}$ as an additional symmetry. Furthermore, this also occurs in the $U \ \rightarrow \infty$ limit and in the trivial $U=0$ case. Thus we conclude that the existence of $P_{\sigma}$ and a careful choice of parameters is what allows us to connect the Hubbard model, which lacks supersymmetry, with the integrable structure of the $AdS_5 \times S^5$ superstring. 

Interestingly, the particle-hole transformation also provides us with a boundary theory which possesses a remnant of the bulk Yangian symmetry. Before constructing the generators of such symmetry, it is necessary to review the theory of twisted Yangians and achiral boundaries. 

\end{section}

\begin{section}{Twisted Yangian symmetry in the presence of achiral boundaries}
Suppose a 1+1D physical theory has a Lie symmetry algebra $\mathfrak{g}$ generated by $Q^a_0$, $a=1,\ldots$,dim($\mathfrak{g}$) satisfying 
\begin{equation}
[Q^a_0,Q^b_0]= f^{ab}_{ \ \ c}Q_0^c.
\end{equation}
For this system to be integrable, it must possess additional conserved charges, and hence it is expected to be invariant under an extension of $\mathfrak{g}$: the Yangian $Y(\mathfrak{g})$ \cite{Mackay}. This is generated by $\{ Q_0^a \}$ - also called the grade-0 generators - and a second set of operators $\{ Q_1^a \}$ which form a vector representation of $\mathfrak{g}$
\begin{equation}
[Q^a_0,Q^b_1]= f^{ab}_{ \ \ c}Q_1^c,
\end{equation} 
and satisfy the so called Drinfel'd relations \cite{Drinfeld}. A commutator of grade-1 generators gives grade-2 generators, and iterating this process one can construct an infinite tower of conserved charges. $Y(\mathfrak{g})$ possesses a coproduct structure, which defines the action of its generators in 2-particle states through the following map
\begin{eqnarray}
\Delta: \ U\mathfrak{g} \ &\rightarrow& \ U\mathfrak{g} \otimes U\mathfrak{g} \nonumber \\
Q_0^a \ &\mapsto& \ Q_0^a \otimes 1+1 \otimes Q_0^a \\
Q_1^a \ &\mapsto& \ Q_1^a \otimes 1+1 \otimes Q_1^a + \frac{1}{2} f^a_{\;\;bc}Q_0^c\otimes Q_0^b.
\end{eqnarray}
where $U\mathfrak{g}$ is the universal enveloping algebra. Finite dimensional representations of $Y(\mathfrak{g})$ are realized in one-parameter families via the automorphism
\begin{eqnarray}
\psi_{\mu}: Y(\mathfrak{g}) \ &\rightarrow& \ Y(\mathfrak{g}) \nonumber \\
(Q^a_0,Q^a_1)\ &\mapsto & \ (Q^a_0,Q^a_1+\mu Q^a_0).
\end{eqnarray}
If a model which possesses $Y(\mathfrak{g})$ is put on the half-line, the boundary condition will break $\mathfrak{g}$ to a subalgebra $\mathfrak{h}$. To determine whether this system possesses a remnant of the original Yangian symmetry referred to as the twisted Yangian $Y(\mathfrak{g},\mathfrak{h})$ \cite{27,DMS,MacKay2}, one must check that $\mathfrak{g}$ and $\mathfrak{h}$ form a symmetric pair : if $\mathfrak{g}=\mathfrak{h}\oplus \mathfrak{m}$, then
\begin{equation}
[\mathfrak{h},\mathfrak{h}]\subset \mathfrak{h}, \quad [\mathfrak{h},\mathfrak{m}]\subset \mathfrak{m}, \quad [\mathfrak{m},\mathfrak{m}]\subset \mathfrak{h}\ ,
\end{equation}
This, together with orthogonality with respect to the killing form of $\mathfrak{g}$, is a requirement for the system to satisfy the coideal property:
\begin{equation}
\Delta Y(\mathfrak{g},\mathfrak{h}) \subset Y(\mathfrak{g}) \otimes Y(\mathfrak{g},\mathfrak{h}).
\end{equation}
Now suppose a 1+1D physical theory has symmetry algebra $\mathfrak{g}_L \times \mathfrak{g}_R$, where $\mathfrak{g}_L$ and $\mathfrak{g}_R$ are generated by $\{J_0^L\}$ and $\{J_0^R\}$ respectively. One can also decompose this symmetry into $\mathfrak{g}_+ \oplus \mathfrak{g}_-$, where $J^{\pm}_0 = J^L_0\pm J^R_0$. If we were to impose an achiral boundary condition on the real line \cite{Vidas3}, which satisfies $\alpha(J_0^L)=J_0^R$ and $\alpha^2=id$, $\mathfrak{g}_L\times \mathfrak{g}_L$ would break to the subalgebra $\mathfrak{g}_+$.

One can check that $\mathfrak{g}_L\times \mathfrak{g}_R$ and $\mathfrak{g}_+$ form a symmetric pair, and hence an integrable system with this type of boundary condition is expected to possess a remnant of the original $Y(\mathfrak{g}\times \mathfrak{g})$ symmetry. This is not $Y(\mathfrak{g})$, but rather, the twisted Yangian $Y(\mathfrak{g}_L\times \mathfrak{g}_R,\mathfrak{g}_+)$ \cite{Vidas3}. Now the task is to construct its generators. It is generated by $\mathfrak{g}_+$ and a deformation of the grade 1 generators $J_1^-=J_1^L-J_1^R$ \cite{vidas3} given by:
\begin{equation}
\widehat{J_1^-}=J_1^-+k[\textbf{C}_+,J_0^-].
\end{equation}
where $k$ is a deformation parameter fixed by the theory and \textbf{C}$_+$ is the quadratic Casimir operator of $\mathfrak{g}$ restricted to $\mathfrak{g}_+$.

\end{section}

\begin{section}{Twisted Yangian of the Hubbard chain with an achiral boundary}
The $\mathfrak{so}(4)$ algebra may be generated by operators $A^a$ and $B^a$, $a=+,-,Z$, satisfying the following relations
\begin{equation}
[A^a,A^b]=f^{ab}_{ \ \ c}A^c, \quad [B^a,B^b]=f^{ab}_{ \ \ c}A^c, \quad 
[A^a,B^b]=f^{ab}_{ \ \ c}B^c,
\end{equation}
where $f^{ab}_{ \ \ c}$ are the $\mathfrak{su}(2)$ structure constants. Note that $\{A^a\}$ generate a full $\mathfrak{su}(2)$ algebra. Since $\mathfrak{so}(4) \cong \mathfrak{su}(2)^2$, $A^a$ and $B^b$ can be constructed via the $\mathfrak{su}(2)_L \times \mathfrak{su}(2)_R$ generators in the following way 
\begin{equation}
A^+_0 = E_{0L}+E_{0R}, \quad A^-_0 = F_{0L}+F_{0R}, \quad A^Z_0 = H_{0L}+H_{0R},
\end{equation}
\begin{equation}
B^+_0 = E_{0L}-E_{0R}, \quad B^-_0 = F_{0L}-F_{0R}, \quad B^Z_0 = H_{0L}-H_{0R}.
\end{equation}
The level 1 generators of the Yangian symmetry are constructed similarly, changing the grade label from 0 to 1. Now consider the following hamiltonian for a half-infinite Hubbard chain:
\begin{equation}
H^{A}=-\sum_{i=-\infty}^{N-1}\sum_{\sigma= \uparrow,\downarrow} c_{i\sigma}^{\dagger}c_{i+1\sigma}+c_{i+1\sigma}^{\dagger}c_{i\sigma}+U\sum_{i=-\infty}^N(n_{i\uparrow}-\frac{1}{2})(n_{i\downarrow}-\frac{1}{2})+pP^+_{N \downarrow}(1),
\end{equation}
The boundary term $pP^+_{N \downarrow}(1)= P^+_{N \downarrow}(p)$ acts on the fundamental states by reflecting a particle with a hole and vice versa at site $N$. In doing this, the states gain a factor of $p$, which is interpreted as a change in phase, requiring $\lvert  p \lvert=1$. Since this specific PHT maps $J_L \in \mathfrak{su}(2)_L$ to $J_R \in \mathfrak{su}(2)_R$ via $P^+_{N \downarrow}(1)J^L(P^+_{N\downarrow}(1))^{-1} = J^R$, this model is no longer invariant under the full $\mathfrak{so}(4)$ algebra, but the symmetry is broken to the diagonal $\mathfrak{su}(2)$ generated by $\{A^a\}$, which we shall denote by $\mathfrak{su}(2)_+$. This is then an achiral boundary condition, and since $\mathfrak{so}(4)$ and $\mathfrak{su}(2)_+$ form a symmetric pair, the model is expected to possess a twisted Yangian symmetry $Y(\mathfrak{so}(4),\mathfrak{su}(2)_+)$. 
Naively, one would attempt to construct the deformed level 1 generators using (4.8) and obtain, for example,
\begin{equation}
\widehat{B}^+ = B_1^+ - \frac{ U}{8}(B_0^+ A_0^Z - B^Z_0 A^+_0),
\end{equation} 
However, just as in the case of other integrable open boundaries \cite{AM}, there exists a subtlety: one must make use of the Yangian automorphism $J_1 \ \mapsto \ J_1+\mu J_0$. In addition, the right Yangian copy is obtained not only by performing the map (2.8) but also by changing $U$ to $-U$. Hence, to satisfy the coideal property, the level 1 generators which must be “twisted” are $A_1^a=J_{1L}^a+J_{1R}^a, a = +,-,Z$. One then finds that the following operator
\begin{equation}
\widetilde{B}^+ = A_1^++\frac{U}{2\sqrt{2}}B_0^+-\frac{U}{4\sqrt{2}}(B_0^+ A_0^Z - B^Z_0 A^+_0)
\end{equation}
commutes with $H^{A}$. Simlarly, the other twisted level 1 charges are:
\begin{eqnarray}
\widetilde{B}^- &=& A_1^--\frac{U}{2\sqrt{2}}B_0^++\frac{U}{4\sqrt{2}}(B_0^- A_0^Z - B^Z_0 A^-_0)\\ \nonumber
\widetilde{B}^Z &=& A_1^Z+\frac{U}{2}B_0^Z+\frac{U}{4}(B_0^+ A_0^- - B^-_0 A_0^+)
\end{eqnarray}

Their coproducts are
\begin{eqnarray}
\Delta \widetilde{B}^+ &=& \widetilde{B}^+ \otimes 1 + 1\otimes \widetilde{B}^+ - \frac{U}{2\sqrt{2}}(B^+_0 \otimes A_0^Z - B^Z_0 \otimes A^+_0) \nonumber \\
\Delta \widetilde{B}^- &=& \widetilde{B}^- \otimes 1 + 1\otimes \widetilde{B}^- + \frac{U}{2\sqrt{2}}(B^-_0 \otimes A_0^Z - B^Z_0 \otimes A^-_0) \nonumber \\
\Delta \widetilde{B}^{Z} &=& \widetilde{B}^Z \otimes 1 + 1\otimes \widetilde{B}^Z + \frac{U}{2}(B^+_0 \otimes A_0^- - B^-_0 \otimes A^+_0) 
\end{eqnarray}
thus (4.7) is satisfied and hence $Y(\mathfrak{so}(4),\mathfrak{su}(2)_+)= \{A^a_0,\widetilde{B}^b \}$ forms a coideal subalgebra of $Y(\mathfrak{so}(4))$.

\end{section}

\begin{section}{Concluding remarks}
In this paper we have shown that the particle-hole transformation plays a crucial role in relating the integrable structure of the Hubbard model to that of the $AdS_5\times S_5$ superstring. Furthermore, we have shown that a particle-hole reflection is an achiral boundary in the half-infinite Hubbard chain, and constructed its corresponding twisted Yangian symmetry. 

These results raise the possibility of studying supersymmetric integrable systems -- especially those relevant in the AdS/CFT correspondence -- using manifestly non-supersymmetric ones. It would be interesting to see if extended Hubbard chains -- possessing an arbitrary symmetry group \cite{FFR} -- or those with variable range hopping \cite{Gohmann} can give rise to interesting integrable boundary theories, and whether these have any relation to other integrable structures in AdS/CFT.

As for the Hubbard model, the tetrahedron algebra is used to obtain the conditions necessary for Shastry's ansatz to satisfy the Yang Baxter equation. The question of whether this algebra is physically relevant in this case remains a mystery. 

\end{section}
\begin{section}{Acknowledgements}
I would like to thank Niall Mackay and Vidas Regelskis for useful discussions and comments on the manuscript. I would also like to thank the Department of Mathematics at the University of York for funding under a studentship.
\end{section}
\begin{section}{Appendix}
\begin{subsection}{The $\mathfrak{su}$(2\textbar 2)$\ltimes \mathbb{R}^2$ relations}
The centrally extended $\mathfrak{su}$(2\textbar 2) superalgebra is generated by six bosonic operators $\{\mathbb{L}^{\alpha}_{ \ \beta},\mathbb{R}^{a}_{ \ b} \}$ and eight supersymmetric generators $\{\mathbb{Q}^{\alpha}_{ \ a},\mathbb{G}^{b}_{ \ \beta} \}$, satisfying the following relations
\begin{align}
&[\mathbb{L}^{\alpha}_{ \ \beta},\mathbb{L}^{\gamma}_{ \ \xi}]=
\delta_{\beta}^{\gamma}\mathbb{L}^{\alpha}_{ \ \xi}-\delta_{\xi}^{\alpha}\mathbb{L}^{\gamma}_{ \ \beta},&
&[\mathbb{R}^{a}_{ \ b},\mathbb{R}^{c}_{ \ d}]=
\delta_b^c\mathbb{R}^{a}_{ \ d}-\delta_d^a\mathbb{R}^{c}_{ \ b}&\nonumber\\ 
&[\mathbb{L}^{\alpha}_{ \ \beta},\mathbb{Q}^{\gamma}_{ \ b}]=
\delta_{\beta}^{\gamma}\mathbb{Q}^{\alpha}_{ \ b}-\frac{1}{2}\delta_{\beta}^{\alpha}\mathbb{Q}^{\gamma}_{ \ b},&
&[\mathbb{L}^{\alpha}_{ \ \beta},\mathbb{G}^{a}_{ \ \gamma}]=
-\delta_{\gamma}^{\alpha}\mathbb{G}^{\alpha}_{ \ \beta}+\frac{1}{2}\delta_{\beta}^{\alpha}\mathbb{G}^{a}_{ \ \gamma}&\nonumber\\
&\{\mathbb{Q}^{\alpha}_{\ a},\mathbb{Q}^{\beta}_{\ b}\}=\epsilon^{\alpha \beta}\epsilon_{ab}\mathbb{P},& &\{\mathbb{G}_{\ \alpha}^{a},\mathbb{G}^{b}_{\ \beta}\}=\epsilon^{ab}\epsilon_{\alpha \beta}\mathbb{K}& \nonumber\\
&\{\mathbb{Q}^{\alpha}_{\ a},\mathbb{G}^{b}_{\ \beta}\}=\delta_a^b\mathbb{L}_{\ \beta}^{\alpha}+\delta_{\beta}^{\alpha}\mathbb{R}_{\ a}^b+\delta_a^b \delta_{\alpha}^{\beta}\mathbb{C}&
\end{align}
where $\mathbb{C}$, $\mathbb{P}$ and $\mathbb{K}$ are central elements. The superalgebra acts on two bosonic $\ket{\phi^a}$ and two fermionic $\ket{\psi^{\alpha}}$ states, $a,\alpha = 1,2$ in the following way:
\begin{align}
&\mathbb{R}^a_{ \ b}\ket{\phi^a}=\delta_b^c\ket{\phi^a}-\frac{1}{2}\delta^a_b\ket{\phi^c}, &&\mathbb{L}^{\alpha}_{ \ \beta}\ket{\psi^{\gamma}}=\delta_{\beta}^{\gamma}\ket{\psi^{\alpha}}-\frac{1}{2}\delta^{\alpha}_{\beta}\ket{\psi^{\gamma}}&\\
&\mathbb{Q}_{ \ a}^{\alpha} \ket{\phi^b} = \textbf{a}\delta^b_a\ket{\psi^{\alpha}}&& \mathbb{Q}_{ \ a}^{\alpha} \ket{\psi^{\beta}} = \textbf{b}\epsilon^{\alpha \beta}\epsilon_{ab}\ket{\phi^b}& \\
&\mathbb{G}_{ \ \alpha}^a \ket{\psi^{\beta}} = \textbf{c}\epsilon^{ab}\epsilon_{\alpha \beta}\ket{\psi^{\beta}}&&\mathbb{G}_{ \ \alpha}^a \ket{\phi^b} = \textbf{d} \delta^{\beta}_{\alpha}\ket{\psi^{\alpha}}& 
\end{align}
where $\textbf{a}$,$\textbf{b}$,$\textbf{c}$ and $\textbf{d}$ are complex numbers and the closure of the algebra requires that $\textbf{ad}-\textbf{bc}=1$, which implies
\begin{equation}
\mathbb{C} = \frac{\textbf{ad}+\textbf{bc}}{2}, \quad \mathbb{P} = \textbf{ab}, \quad \mathbb{K} = \textbf{cd}
\end{equation}
\end{subsection}
\begin{subsection}{Commutation with the Hamiltonian}
We will proceed to show that $\widetilde{B}^+$ commutes with the achiral Hamiltonian $H^A$. Here we will construct the half-infinite Hubbard chain by folding an infinite one at a spin site, say $N$, and identifying sites $N+n$ and $N-n$. Such identification commutes with the particle-hole transformation. Hence, since all components of $\widetilde{B}^+$ are already conserved charges of an infinite Hubbard chain, we only need to show that 
\begin{equation}
[P_{N \downarrow}^{+},\widetilde{B}^+]=0
\end{equation}
It is helpful to divide the commutator into components.  First, let us compute $[P^+_{N \downarrow }, A^+_1]$ by dividing $A^+_1$ into an $U$-independent and dependent components $A^{+0}_1$ and $A^{+U}_1$. For the commutator with $A^{+0}_1$, it is convenient to write $P^+_{N \downarrow}$ in the fermionic representation:
\begin{equation}
P_{N\downarrow}^+=c^{\dagger}_{N\downarrow}-(-1)^N c_{N\downarrow}.
\end{equation}
 Then we find that 
\begin{eqnarray}
[P^+_{N \downarrow }, A^{+0}_1] &=& [c^{\dagger}_{N\downarrow}-(-1)^N c_{N\downarrow}, c_{N-1\uparrow}^{\dagger}c_{N\downarrow}-c_{N\uparrow}^{\dagger}c_{N-1\downarrow}+(-1)^N(c_{N-1\uparrow}^{\dagger}c_{N\downarrow}^{\dagger}+c_{N\uparrow}^{\dagger}c_{N-1\downarrow}^{\dagger} )\nonumber \\
&=& -(c_{N-1 \uparrow}^{\dagger} - (-1)^{2N} c_{N-1 \uparrow}^{\dagger}) \nonumber \\
&=& 0
\end{eqnarray}
For $A^{+U}_2$, as we will see, it is not necessary to compute the commutator of $P^+_{N \downarrow }$ with the different operators, but rather it is sufficient to know that $[P^+_{N \downarrow }, B^+_0] = 2 [P^+_{N \downarrow }, E_{0L}]$, which can be inferred by the relation $(P^+_{N \downarrow })^{-1} E_{0R}P^+_{N \downarrow } = E_{0L}$. We find that 
\begin{equation}
[P^+_{N \downarrow },A^{+U}_2 ]= \frac{U}{2 \sqrt{2}}\sum_{i<N}([P^+_{N \downarrow }, \mathcal{E}^0_{NL}](\mathcal{H}_{iL}^0+\mathcal{H}_{iR}^0)-[P^+_{N \downarrow }, \mathcal{H}^0_{NL}](\mathcal{E}_{iL}^0+\mathcal{E}_{iR}^{0}))
\end{equation}
If one makes the ansatz that the quadratic modification must be of the form $X_B^+ = \mu B_0^+-k(B_0^+ A_0^Z - B^Z_0 A^+_0)$, then
\begin{equation}
[P^+_{N \downarrow }, X_B^+] = -(4k+2\mu)[P^+_{N \downarrow }, \mathcal{E}^0_{NL}]+2k \sum_{i<N}([P^+_{N \downarrow }, \mathcal{E}^0_{NL}](\mathcal{H}_{iL}^0+\mathcal{H}_{iR}^0)-[P^+_{N \downarrow }, \mathcal{H}^0_{NL}](\mathcal{E}_{iL}^0+\mathcal{E}_{iR}^{0}))
\end{equation}
Hence we arrive at the conclusion that $[P^+_{N \downarrow }, A^+_1+X_B^+]=0$ if
\begin{equation}
 k = - \frac{U}{4\sqrt{2}}, \quad \mu = \frac{U}{2\sqrt{2}}. 
 \end{equation}

\end{subsection}

\begin{subsection}{Computation of coproducts}
We will proceed to compute $\Delta \widetilde{B}^+$. Defining $B^+_0 = \sum_i B^{+ (0)}_i$ where $B_{i}^{+ (n)}= \mathcal{E}_{iL}^n-\mathcal{E}_{iR}^n$ and $A^+_0 = \sum_i A^{+ (0)}_i$ where $A_{i}^{+ (n)}= \mathcal{E}_{iL}^n+\mathcal{E}_{iR}^n$ we can rewrite 
\begin{eqnarray}
A^+_1 &=& \frac{1}{\sqrt{2}}\sum_i (\mathcal{E}_{iL}^1-\mathcal{E}_{iL}^{-1}+\mathcal{E}_{iR}^1-\mathcal{E}_{iR}^{-1})-\frac{U}{2\sqrt{2}}\sum_{i,j}t_{ij}(\mathcal{E}_{iL}^0 \mathcal{H}_{jL}^0-\mathcal{E}_{iR}^0 \mathcal{H}_{jR}^0) \nonumber \\
&=& \sum_i (A^{+ (1)}_{i}+A^{+ (-1)}_{i})-\frac{U}{4\sqrt{2}}\sum_{i,j} t_{ij}(B^{+ (0)}_{i}A^{z (0)}_{j}-A^{+ (0)}_{i}B^{z (0)}_{j})
\end{eqnarray}
where $t_{ij}$ is 1 when $j>i$, $-1$ when $j<i$ and 0 when $i=j$. Using the appendix A in \cite{AM}, one can show that
\begin{equation}
\Delta A_1^+ = A_1^+ \otimes 1+1 \otimes A_1^+- \frac{U}{4\sqrt{2}}(B_0^+ \otimes A^z_0+A^+_0 \otimes B_0^z - A^z_0 \otimes B^+_0 - B^z_0 \otimes A^+_0)
\end{equation}
Since $\Delta$ is a homomorphism,
\begin{eqnarray}
\Delta \widetilde{B}^+ &=& \Delta A_1^+-\frac{U}{2\sqrt{2}} \Delta B_0^+-\frac{U}{4\sqrt{2}}(\Delta B_0^+ \Delta A_0^Z - \Delta B^Z_0 \Delta A^+_0) \nonumber \\
&=&  A_1^+ \otimes 1 + 1 \otimes A_1^+ -\frac{U}{4\sqrt{2}}(2B_0^+ \otimes A^z_0 - 2B_0^z \otimes A^+_0 + (B_0^+A^z_0-B^+_0A^z_0)\otimes 1 \nonumber \\
&& +1 \otimes(B_0^+A^z_0-B^+_0A^z_0) )- \frac{U}{2\sqrt{2}}(B_0^+ \otimes 1 + 1 \otimes B_0^+) \nonumber \\
&=& \widetilde{B}^+ \otimes 1 + 1\otimes \widetilde{B}^+ - \frac{U}{2\sqrt{2}}(B^+_0 \otimes A_0^Z - B^Z_0 \otimes A^+_0)
\end{eqnarray}
\end{subsection}
\end{section}

\end{document}